\documentstyle[12pt]{article}
\begin{document}
\begin{center}
{\bf Ricci Collineations of the Bianchi Type-II, VIII,
and IX Spacetimes}\\[1cm]
\.{I}. Yavuz$^{\dagger}$ and U. Camc{\i}$^{\dagger}$\\[1cm]
$^\dagger$ \d{C}anakkale Onsekiz Mart University, Art and Science Faculty,
Department of Mathematics, 17100 \d{C}anakkale, Turkey \\
e-mail : iyavuz@comu.edu.tr, \\
e-mail : ucamci@hotmail.com
\end{center}
\begin{abstract}
Ricci and contracted Ricci collineations of the Bianchi type II, VIII, and
IX space-times, associated with the vector fields of the form (i) one
component of $\xi^a(x^b)$ is different from zero and (ii) two components of
$\xi^a(x^b)$ are different from zero, for $a,b =1,2,3,4,$ are presented. In
subcase (i.b), which is $\xi^a = (0,\xi^2(x^a),0,0)$, some known solutions
are found, and in subcase (i.d), which is $\xi^a=(0,0,0,\xi^4(x^a))$,
choosing $S(t)=const.\times R(t)$, the Bianchi type II, VIII, and IX
spacetimes is reduced to the Robertson-Walker metric.
\end{abstract}

\section{Introduction}

      In addition to isometries that leave invariant the metric tensor under
Lie transport and provide information of the symmetries inherent in
space-time, spacetimes may admit other symmetries that do not leave
the metric tensor invariant. Whereas the geometrical nature of
a space-time is encoded into the metric tensor, by virtue of the Einstein
equations, the physics is given more explicitly by the Ricci tensor.
Besides the space-time metric, the curvature and Ricci tensors are other
important candidates which play the significant role in understanding
the geometric structure of space-times in general relativity. These general
symmetry considerations (e.g., Ricci collineations, curvature collineations),
based on the invariance properties of different geometric objects under Lie
transport, have been classified by Katzin et al.\cite{Katzin}

       The locally rotationally symmetric (LRS) metric for the spatially
homogeneous  Bianchi type-II($\delta=0$), VIII($\delta=-1$), and
IX($\delta=+1$) cosmological models can generally be written in the
form\cite{Banerjee1,Sing}
\begin{equation}
ds^2 = dt^2-S^2(dx-hdz)^2-R^2(dy^2 + f^2 dz^2)
\end{equation}
where $S$ and $R$ are arbitrary functions of  $t$ only and
\begin{eqnarray}
f(y)=\left(\begin{array}{c}
y \\ siny \\ sinhy
\end{array} \right),
h(y)=\left(\begin{array}{c}
-y^2/2 \\ cosy \\ -coshy
\end{array} \right)\,\,\,{\small for}\,\,\,
\delta = -\frac{f^{\prime \prime}}{f} = \left(\begin{array}{c}
0 \\ +1 \\ -1
\end{array} \right) \nonumber
\end{eqnarray}
(prime denotes differentiation with respect to $y$).

       We give the geometric classification of 3-parameter Lie groups
for the Bianchi type II, VIII, and IX space-times in the following table
(from Ref.$4$). In this table, the last column gives a possible realisation
of the algebra in terms of translations, rotations, boosts, null rotations
and a dilation (scaling) acting on subspaces of Minkowski space, where $a$
and $b$ are the parameters which occur in the generators ($a=0$ for Bianchi
type V, $a=b$ for Bianchi type IV).
\\
   {\bf Table I.}Geometric classification of 3-parameter Lie groups for the
      Bianchi type II, VIII, and IX space-times.\\
\begin{tabular}{|p{0.4in}|p{0.7in}|p{1.4in}|c|} \hline
Group type  & Isomorphic to  &  (Sub)group of & Generators \\ \hline
II   & $G_1, H_1$ & homothetic motions of null $2$-plane & $\partial_y,\partial_x, a\partial_y+b(x\partial_x+y\partial_y)$ \\
VIII & $SO(2,1)$  & $2$-pseudosphere  & $t\partial_x+x\partial_t, t\partial_y+y\partial_t, x\partial_y-y\partial_x$  \\
IX   & $SO(3)$    & $2$-sphere        & $y\partial_x-x\partial_y, z\partial_y-y\partial_z, x\partial_z-z\partial_x$ \\ \hline
\end{tabular}
\\

     The Killing equation can be written in component form as
\begin{equation}
(K_{ab}) : g_{ab,c}k^c + g_{ac}k^c_{,b} + g_{cb}k^c_{,a} = 0,
\end{equation}
where the comma represents partial differentiation with respect to the
respective coordinate, $k^a$ are the components of the KV field, and
Latin indices take values $1,2,3,4$. As for all Bianchi space-times, the
Bianchi type II, VIII, and IX space-times considered here also admit Killing
Vectors (KVs), or isometries \cite{Kramer} and homothetic motions \cite{Banerjee2},
i.e. symmetries that scale all distances by the same constant factor and
preserve the null geodesic affine parameters.

     According to the classification of Katzin et al. \cite{Katzin},
a given Riemannian space-time will admit Ricci collineations (RCs) provided
one can find a solution of the equations \cite{Davis,Nunez}
\begin{equation}
\pounds_\xi R_{ab}=\xi^c \nabla_c R_{ab} + R_{ac}\nabla_b\xi^c
+R_{cb}\nabla_a\xi^c = 0,
\end{equation}
where $\pounds_\xi$ denotes the operation of Lie differentiation with respect
to the vector $\xi^a$ and $\nabla$ represents covariant derivative operator.
Clearly, if a solution $\xi^a$ to equation (3) exists (which corresponds to an
infinitesimal point mapping $x^a\rightarrow x^a + \epsilon \xi^a$), then it
represents symmetry property of the particular Riemannian spacetime, and
this symmetry property will correspond to at least a $G_1$ group of Ricci
collineation. In a torsion free space in a coordinate basis, the RC equation
(3) represented by $(C_{ab})$ reduces to a simple partial differential
equation (pde)\cite{Bokhari}
\begin{equation}
(C_{ab}) : R_{ab,c}\xi^c + R_{ac}\xi^c_{,b} + R_{cb}\xi^c_{,a}=0,
\end{equation}
where $\xi^a$ are the components of  the RC vector. Also, a less restrictive
class of symmetries corresponds to the Family of Contracted Ricci
Collineation(FCRC), defined by
\begin{equation}
g_{ab}\pounds_\xi R_{ab}=0.
\end{equation}
Each member of the FCRC symmetry mapping satisfies $g^{ab}\pounds_\xi(T_{ab}
-\frac{1}{2}T g_{ab})=0$, which leads to the following conservation law
generator for a space-time admitting RC \cite{Nunez}:
\begin{eqnarray}
\nabla_a[(-g)^{\frac{1}{2}}(T^a_b-\frac{1}{2}T\delta^a_b)\xi^b] = \nabla_a[
(-g)^{\frac{1}{2}}R^a_b\xi^b] = \partial_a[(-g)^{\frac{1}{2}}R^a_b\xi^b]=0. \nonumber
\end{eqnarray}

        The relationship between the RCs and the isometries have been
discussed in detail by Amir, Bokhari, and Qadir \cite{Amir}. They provide
the complete classification of the RCs according to the nature of the
Ricci tensor which is constructed from a general spherically symmetric and
static metric.
\\

        If $h=0$ and $f=1$, then the metric (1) coincide the LRS Bianchi
type I space-time \cite{MacCallum} while if we take $h=0, f=1$ and  $R=S$ in
the metric (1), then we have the Robertson-Walker spacetime with $k=0$.
Green et al.\cite{Green} and Nu$\tilde{n}$ez et al.\cite{Nunez} have provided an
example of RC and FCRC symmetries of  Robertson-Walker spacetime, and they
have confined their study to symmetries generated by the vector fields of
the form, respectively,
\begin{equation}
\xi =(0,0,0,\xi^4(r,\theta,\phi,t)),
\end{equation}
and
\begin{equation}
\xi =(\xi^1(r,t),0,0,\xi^4(r,t)).
\end{equation}
In this paper we investigate some symmetry properties of the Bianchi type II,
VIII, and IX space-times by considering RCs associated with the following
vector fields \\
(i) one components of $\xi^a(x^b)$ is different from zero:
\begin{eqnarray}
(i.a)\,\,\,\, \xi^a &=&(\xi^1(x^a),0,0,0), \nonumber \\
(i.b)\,\,\,\, \xi^a &=&(0,\xi^2(x^a),0,0), \nonumber \\
(i.c)\,\,\,\, \xi^a &=&(0,0,\xi^3(x^a),0), \nonumber \\
(i.d)\,\,\,\, \xi^a &=&(0,0,0,\xi^4(x^a)), \nonumber
\end{eqnarray}
(ii) two components of $\xi^a(x^b)$ is different from zero. In this case, we
consider the following subcases only.
\begin{eqnarray}
(ii.a)\,\,\,\, \xi^a &=&(\xi^1(x^a),\xi^2(x^a),0,0), \nonumber \\
(ii.b)\,\,\,\, \xi^a &=&(0,\xi^2(x^a),\xi^3(x^a),0), \nonumber \\
(ii.c)\,\,\,\, \xi^a &=&(0,\xi^2(x^a),0,\xi^4(x^a)), \nonumber
\end{eqnarray}
where $x^a=(x,y,z,t)$. Furthermore, for these vector fields, we obtain a FCRC
symmetry family. The motivation to study this kind of symmetries comes from
the fact that, to the best of our knowledge, a systematic analysis of RCs and
FCRC associated to Bianchi universes is not available in the literature.

      In section 2, we study in detail RCs for the vector fields in the cases
(i) and (ii) and also, in section 3, we are dealt with FCRC.

\section{Ricci Collineations}

     The nonvanishing components of the Ricci tensor, corresponding to
the metric (1), read
\begin{eqnarray}
R_{11}&=&S^2A(t),\,\,\,R_{13}=-h[R_{11}+\beta S^2/(2R)^2],\,\,\,R_{22}=R^2 B(t), \nonumber \\
\\
R_{33}&=&h^2[R_{11}+\beta S^2/R^2] + f^2 R_{22},\,\,\,R_{44}=(\frac{2\ddot{R}}{R}
+\frac{\ddot{S}}{S}) \nonumber
\end{eqnarray}
where a dot indicates derivative with respect to time and we have  defined
\begin{equation}
A(t)=\frac{\ddot{S}}{S} + 2\frac{\dot{R} \dot{S}}{RS} + \alpha\frac{S^2}{2R^4},\,\, 
B(t)=\frac{\ddot{R}}{R} + \frac{\dot{R} \dot{S}}{RS} + \left(\frac{\dot{R}}{R}
\right)^2 - \left[\alpha \frac{S^2}{2R^4} - \frac{\delta}{R^2} \right] 
\end{equation}
such that $\alpha$ and $\beta$ are
\begin{eqnarray}
\beta &=& \frac{f' h'}{f\,h}-\frac{h''}{h},\,\,\,\,\,\,
\alpha = \left( \frac{h'}{f} \right)^2. \nonumber
\end{eqnarray}
For the Bianchi type-II, VIII, and IX spacetime, we obtain that $\alpha=1$
and $\beta=0$.

   For the metric (1), the RC equations (4), generated by an
arbitrary vector field $\xi^a(x,y,z,t)$, reads \\ \\
$(C_{11}): R_{11,4}\xi^4 + 2 R_{11}\xi^1_{,1} + 2 R_{13}\xi^3_{,1}=0,$ \\ \\
$(C_{22}): R_{22,4}\xi^4 + 2 R_{22}\xi^2_{,2}=0,$ \\ \\
$(C_{33}): R_{33,2}\xi^2 + R_{33,4}\xi^4 + 2 R_{13}\xi^1_{,3} + 2 R_{33}\xi^3_{,3}=0,$ \\ \\
$(C_{44}): R_{44,4}\xi^4 + 2 R_{44}\xi^4_{,4}=0,$ \\ \\
$(C_{12}): R_{11}\xi^1_{,2} + R_{22}\xi^2_{,1} + R_{13}\xi^3_{,2}=0,$ \\ \\
$(C_{13}): R_{13,2}\xi^2 + R_{13,4}\xi^4 + R_{11}\xi^1_{,3} + R_{33}\xi^3_{,1}
+ R_{13}(\xi^1_{,1} + \xi^3_{,3})=0,$ \\ \\
$(C_{14}): R_{11}\xi^1_{,4} + R_{44}\xi^4_{,1} + R_{13}\xi^3_{,4}=0,$ \\ \\
$(C_{23}): R_{22}\xi^2_{,3} + R_{33}\xi^3_{,2} + R_{13}\xi^1_{,2}= 0,$ \\ \\ 
$(C_{24}): R_{22}\xi^2_{,4} + R_{44}\xi^4_{,2}=0,$ \\ \\
$(C_{34}): R_{33}\xi^3_{,4} + R_{44}\xi^4_{,3} + R_{13}\xi^1_{,4}=0.$ \\ \\
\\
In the above equations we will consider two different cases : \\ \\
{\bf Case (i):} One component of $\xi^a(x^b)$ is different from zero. \\
In this case, there are four subcases: \\ \\
{\it Subcases (i.a)-(i.c)}. From the RC equations, for these three subcases,
we obtain that $\xi^1, \xi^2$ and $\xi^3$ are constants. It means that these
three RC vectors $\xi^i, i=1,2,3$ represent a {\it translation} along $x,
y$ and $z$ directions, respectively. Further, from subcase (i.b) we get
\begin{equation}
R_{11}= R_{22}=0\,\,\,\,\, \Leftrightarrow \,\,\,\,\, A(t) = B(t) = 0.
\end{equation}
Therefore, using (8), (9) and (10), we find that
\begin{eqnarray}
\frac{\ddot{R}}{R}+\frac{\ddot{S}}{S}+3\frac{\dot{R}\dot{S}}{RS}+
(\frac{\dot{R}}{R})^2 + \frac{\delta}{R^2} &=& 0, \nonumber
\end{eqnarray}
which leads to
\begin{equation}
[R (RS)^{\dot{}}\,\,]^{\dot{}} = -\delta S.
\end{equation}
When we use the transformations of the time-coordinate $d\bar{t} = Sdt$ and
$dt=SR^2d\tau$ in (11), we obtain the general solutions for Bianchi type
II($\delta=0$) metric, respectively,
\begin{equation}
(RS)^2 =c_1 \bar t + c_2,\,\,\,\,{\small and}\,\,\,\,(RS)^2 =e^{2(c_3\tau + c_4)},
\end{equation}
where $c_1, c_2, c_3,$ and $c_4$ are integration constants. These solutions
are found by Lorenz \cite{Lorenz1}. Also, making the scale transformation
$d\bar{t} = Sdt$ into (11), we get the general solution for Bianchi type II,
VIII, and IX metrics,
\begin{equation}
(RS)^2 = -\delta (\bar{t})^2 + c_5\bar{t} + c_6,
\end{equation}
where $c_5, c_6$ are constants. This corresponds to the solution found by
Lorenz \cite{Lorenz2}.\\ \\
{\it subcase (i.d)}. From the RC equations, we get \\ \\
$(C_{11}): R_{11,4}\xi^4 = 0,\,\,\,(C_{22}): R_{22,4}\xi^4 = 0,\,\,\,(C_{33}): R_{33,4}\xi^4 = 0,$ \\ \\
$(C_{44}): R_{44,4}\xi^4 + 2 R_{44}\xi^4_{,4} = 0,$ \\ \\
$(C_{13}): R_{13,4}\xi^4 = 0,\,\,\,\,(C_{14}): R_{44}\xi^4_{,1} = 0,$ \\ \\
$(C_{24}): R_{44}\xi^4_{,2} = 0,\,\,\,(C_{34}): R_{44}\xi^4_{,3} = 0.$ \\ \\
In this case, from $(C_{11})$ and $(C_{22})$, we find that
\begin{equation}
\dot{R}_{11} = 0 \,\,\,\,\,\Leftrightarrow \,\,\,\,\, R_{11}=a_1,
\end{equation}
\begin{equation}
\dot{R}_{22} = 0 \,\,\,\,\,\Leftrightarrow \,\,\,\,\, R_{22}=a_2,
\end{equation}
where $a_1, a_2$ are constants. It is clear that Eqs. $(C_{33})$ and
$(C_{13})$ are identically satisfied. Further, from $(C_{14}), (C_{24})$ and
$(C_{34})$, we have $\xi^4=\xi^4(t)$. Now, using $(C_{44})$, we obtain that
\begin{equation}
\xi^4(t) =\frac{c}{\sqrt{\mid{R_{44}\mid{}}}},
\end{equation}
where $c$ is a constant. Choosing $S(t) = a_3 R(t), a_3$ = const., from (14)
and (15) we obtain
\begin{equation}
\frac{\ddot{R}}{R} + 2\left(\frac{\dot{R}}{R} \right)^2 + \frac{2k}{R^2} =0,
\end{equation}
where $k$ is a disposable constant, and can be set to $\pm1$ or $0$. Then,
the solution for eq. (17) represents the Robertson-Walker spacetime in
general. The RCs of this space-time for the RC vector (7) and the solutions
for eq. (17) have been found by Nu$\tilde{n}$ez et al. \cite{Nunez} \\ \\
{\bf Case (ii)}. The two components of $\xi^a(x^b)$ are different from zero.
In this case, we will consider the following subcases only. \\ \\
{\it Subcase (ii.a)}. From the RC equations, we obtain that \\ \\
$(C_{33}): R_{33,2}\xi^2 + 2 R_{13}\xi^1_{,3} = 0,$ \\ \\
$(C_{12}): R_{11}\xi^1_{,2} + R_{22}\xi^2_{,1} = 0,$ \\ \\
$(C_{13}): R_{11}\xi^1_{,3} + R_{13,2}\xi^2 = 0,$ \\ \\
$(C_{23}): R_{13}\xi^1_{,2} + R_{22}\xi^2_{,3} = 0,$ \\ \\
$(C_{11}): R_{11}\xi^1_{,1} = 0,\,\,\,(C_{22}): R_{22}\xi^2_{,2} = 0, $ \\ \\
$(C_{14}): R_{11}\xi^1_{,4} = 0,\,\,\,(C_{24}): R_{22}\xi^2_{,4} = 0. $ \\ \\
From the above equations, using $(C_{11}), (C_{22}), (C_{14})$ and $(C_{24})$,
we find that $\xi^1 = \xi^1(y,z)$ and $\xi^2=\xi^2(x,z)$. As a result of
$(C_{13})$ and $(C_{23})$, we obtain that $\xi^2$ vanishes, and therefore
$\xi^1$ becomes a constant. Then it follows that there is a {\it translation}
along the $x$-direction, since $\xi^2$ vanishes and $\xi^1$ is a constant. \\ \\
{\it Subcase (ii.b)}. For this subcase, RC equations yield \\ \\
$(C_{33}): R_{33,2}\xi^2 + 2 R_{33}\xi^3_{,3}= 0,$ \\ \\
$(C_{12}): R_{22}\xi^2_{,1} + R_{13}\xi^3_{,1} = 0,$ \\ \\
$(C_{13}): R_{33}\xi^3_{,1} + R_{13}\xi^3_{,3} = 0,$ \\ \\
$(C_{23}): R_{22}\xi^2_{,3} + R_{33}\xi^3_{,2} = 0,$ \\ \\
$(C_{11}): 2R_{13}\xi^3_{,1} = 0,\,\,\,\,(C_{22}): 2R_{22}\xi^2_{,2} = 0,$ \\ \\
$(C_{14}): R_{13}\xi^3_{,4} = 0,\,\,\,\,\,(C_{24}):  R_{22}\xi^2_{,4} = 0.$ \\ \\
Using $(C_{22})$ and $(C_{24})$, we have $\xi^2 = \xi^2(x,z)$, while
eqs. $(C_{11})$ and $(C_{14})$ give $\xi^3=\xi^3(x,z)$. Therefore, eq.
$(C_{13})$ becomes 
\begin{eqnarray}
(C_{13}): R_{13}\xi^3_{,3} + R_{13,2}\xi^2 = 0. \nonumber
\end{eqnarray}
From this equation, obviously, we find
\begin{equation}
\xi^2(x,z) = -\frac{h}{h'} \xi^3_{,3}.
\end{equation}
Then, substituting eq. (18) into $(C_{33})$, we obtain that $\xi^3$ is a
function of $y$ only. Therefore, from (18), we find that $\xi^2$ vanishes and
also, from $(C_{12})$ and $(C_{23})$, that $\xi^3$ is a constant. In this
case, since $\xi^2$ vanishes and $\xi^3$ is a constant, this means that
there is a {\it translation} along the z-direction.\\ \\
{\it Subcase (ii.c)}. In this subcase, we have the following RC equations: \\
\\
$(C_{11}): R_{11,4}\xi^4 = 0,$ \\ \\
$(C_{22}): R_{22,4}\xi^4 + 2 R_{22}\xi^2_{,2} = 0,$ \\ \\
$(C_{33}): R_{33,2}\xi^2 + R_{33,4}\xi^4 = 0,$ \\ \\
$(C_{44}): R_{44,4}\xi^4 + 2 R_{44}\xi^4_{,4} = 0,$ \\ \\
$(C_{13}): R_{13,2}\xi^2 + R_{13,4}\xi^4 = 0,$ \\ \\
$(C_{24}): R_{22}\xi^2_{,4} + R_{44}\xi^4_{,2} = 0,$ \\ \\
$(C_{12}): R_{22}\xi^2_{,1} = 0,\,\,\,\,(C_{14}): R_{44}\xi^4_{,1} = 0,$ \\ \\
$(C_{23}): R_{22}\xi^2_{,3} = 0,\,\,\,\,(C_{34}): R_{44}\xi^4_{,3} = 0,$ \\ \\
Using eqs. $(C_{12}), (C_{14},(C_{23})$ and $(C_{34})$, we have that
$\xi^2$ and $\xi^4$ are functions of $y$ and $t$ only. From $(C_{11})$, we
find that
\begin{eqnarray}
\dot{R}_{11} &=& 0\,\,\,\,\, \Leftrightarrow\,\,\,\,\, R_{11} = const. =a_1 \nonumber
\end{eqnarray}
and by using $(C_{13})$ we see that $\xi^2$ vanishes. Therefore, eqs. $(C_{22})$
and $(C_{33})$ give
\begin{eqnarray}
\dot{R}_{22} = 0\,\,\,\,\, \Leftrightarrow\,\,\,\,\, R_{22} = const. =a_2 \nonumber
\end{eqnarray}
On the other hand, the component $\xi^4$ is determined by $(C_{44})$, and
since from $(C_{24})$ that $\xi^4=\xi^4(t)$, we see that it keeps the form
(16). Therefore, this subcase is reduced to the subcase (i.d).

\section{Family of Contracted Ricci Collineations}

      The Family of Contracted Ricci collineations (FCRC) for the Bianchi
type II, VIII, and IX metric (1) takes the form
\begin{eqnarray}
A(t)\xi^1_{,1} + B(t)\left(\frac{f'}{f}\xi^2 + \xi^2_{,2} \right) +
h[B(t)-A(t)]\xi^3_{,1} \nonumber\\
+ B(t)\xi^3_{,3} + R_{44} \left[ \left(
\frac{\dot{R}_{44}}{R_{44}} + 2\frac{\dot{R}}{R} + \frac{\dot{S}}{S} \right)
\xi^4 + 2 \xi^4_{,4} \right] &=& 0.
\end{eqnarray}
It is not possible to find a solution to (19) without imposing some
additional restrictions either on the metric or on the vector field $\xi^a$.

        In subcase (i.a), we find from (19) that $\xi^1_{,1} = 0,$ i.e.
$\xi^1 = \xi^1(y,z,t)$ for $B(t) \neq 0$. A first example of proper
(nondegenerate) FCRC vector is obtained by setting subcase (i.b) in the above
equation. In the present case, eq.(19) becomes
\begin{eqnarray}
B(t)\left( \frac{f'}{f}\xi^2 + \xi^2_{,2} \right) &=& 0. \nonumber
\end{eqnarray}
This equation can be integrated by demanding that $B(t)$ be different from zero.
Thus we find
\begin{eqnarray}
\xi^2 &=& \frac{k_1(x,z,t)}{f(y)}, \nonumber
\end{eqnarray}
where $k_1(x,z,t)$ is an arbitrary function of integration with respect to
the coordinate $y$ and $f(y) = y, siny,$ or $sinhy$ for the Bianchi types II,
VIII, and IX respectively. By inserting subcase (i.d) in (19), we get
\begin{eqnarray}
\left( \frac{\dot{R}_{44}}{R_{44}} + 2\frac{\dot{R}}{R} + \frac{\dot{S}}{S}
\right) \xi^4 + 2 \xi^4_{,4} &=& 0. \nonumber
\end{eqnarray}
If we solve this equation, we can easily obtain
\begin{eqnarray}
\xi^4 &=& \left( \frac{k_2(x,y,z)}{R^2 S R_{44}} \right)^{1/2} \nonumber
\end{eqnarray}
where $k_2(x,y,t)$ is arbitrary function of the spatial variables.

\section{Conclusion}

     RC equation (4) for metric (1) represents ten nonlinear pde's for seven
unknowns (four $\xi^a, a=1,2,3,4$, which are functions of all the space-time
coordinates, three functions of $t$ given by $R_{11}, R_{22}$, and $R_{44}$,
since $R_{13}=hR_{11}$ and $R_{33} = h^2R_{11} + f^2R_{22}$). In the present
paper, the possible value of the RC classification of Bianchi type II, VIII,
and IX space-times for the vector fields (i) and (ii) are given.

     In section $2$,  we worked out a classification of the RCs for the
RC vector of the form (i) and (ii). In the subcases (i.a), (i.b) and
(i.c), we find from the RC equations $(C_{ab})$ that the components $\xi^1,
\xi^2$ and $\xi^3$ are constants for the Bianchi type II, VIII, and IX
space-time (1). Therefore, it follows that these three subcases represent
{\it translation} along $x,y$ and $z$ directions, respectively. Further, when
we choose RC vector $\xi^a$, as for subcases (ii.a), (ii.b) and (ii.c),
then $\xi^2$ always vanishes and the other two components (i.e., $\xi^1$ and
$\xi^3$) become a constant, but $\xi^4$ is a function of $t$ only. Therefore,
the subcases (ii.a), (ii.b) and (ii.c) are reduced to the subcases (i.a),
(i.c) and (i.d), respectively. Consequently, it should be noted that (i.d)
and (ii.c) types of symmetry vector $\xi^a$ would lead to proper RC for this
cosmological model; the other cases are degenerate (i.e. nonproper). Notice
that every KV is an RC vector for the LRS Bianchi type II, VIII, and IX
space-times, but the converse is not true. We remark that solutions (12)
and (13) which were found by Lorenz \cite{Lorenz1,Lorenz2} for metric (1)
have been obtained by us in subcase (i.b). Also, in subcase (i.d), our metric
(1) is reduced to the Robertson-Walker space-time as a special case if we
choose $S(t)=const.\times R(t)$. Finally, in section 3, for the subcases
(i.b) and (i.d), we find some proper FCRC vector.

        The RC equations $(C_{ab})$ have been abtained by assuming that
$A(t), B(t)$ [i.e.,$R_{11}, R_{13}, R_{22},$ and $R_{33}$], and $R_{44}$
do not vanish. Nevetheless there may exist solutions that correspond to
the vanishing of any of the above quantities. This case will be discussed
in another paper.

\end{document}